\documentclass[superscriptaddress,pre,aps,twocolumn,showpacs,floatfix]{revtex4}

\usepackage[dvips]{graphicx}
\usepackage{amsmath}
\usepackage{amssymb}
\usepackage[nice]{nicefrac}
\usepackage{color}

\newcommand{\xt}{x_{\mathrm{th}}}
\newcommand{\so}{\sigma_{\mathrm{opt}}}
\newcommand{\etamax}{\eta_{\mathrm{max}}}

\begin{document}
\bibliographystyle{prsty}
\title{Non-Gaussian, non-dynamical stochastic resonance}

\author{Krzysztof Szczepaniec}
\email{kszczepaniec@th.if.uj.edu.pl}
\affiliation{Marian Smoluchowski Institute of Physics, and Mark Kac Center for Complex Systems Research, Jagiellonian University, ul. Reymonta 4, 30--059 Krak\'ow, Poland }

\author{Bart{\l}omiej Dybiec}
\email{bartek@th.if.uj.edu.pl}
\affiliation{Marian Smoluchowski Institute of Physics, and Mark Kac Center for Complex Systems Research, Jagiellonian University, ul. Reymonta 4, 30--059 Krak\'ow, Poland }

\date{\today}
\begin{abstract}
The archetypal system demonstrating stochastic resonance is nothing more than a threshold triggered device. It consists of a periodic modulated input and noise. Every time an output crosses the threshold the signal is recorded. Such a digitally filtered signal is sensitive to the noise intensity. There exist the optimal value of the noise intensity resulting in the ``most'' periodic output.  Here, we explore properties of the non-dynamical stochastic resonance in non-equilibrium situations, i.e. when the Gaussian noise is replaced by an $\alpha$-stable noise. We demonstrate that non-equilibrium $\alpha$-stable noises, depending on noise parameters, can either weaken or enhance the non-dynamical stochastic resonance.
\end{abstract}

 \maketitle

\section{Introduction\label{sec:introduction}}

Stochastic resonance \cite{benzi1981,mcnamara1989,gammaitoni1998,anishchenko1999} is one of effects demonstrating constructive role of noises in physical systems. In the stochastic resonance,  a weak input signal, due to presence of a stochastic component in the system dynamics, is amplified and consequently detectable. The presence of the stochastic resonance is an universal feature of barrier crossing events over a periodically modulated potential  \cite{dykman1992,pankratov2000,dybiec2009}. The seminal system demonstrating the stochastic resonance is an overdamped Brownian particle moving in a double well, fourth order periodically modulated potential. The joint action of a periodic modulation and an optimal level of noise result in periodic character of transitions of a test particle over the potential barrier.  Consequently, a weak periodic signal (periodic barrier modulation) due to the presence of a noise is amplified and detectable. An analysis of the stochastic resonance \cite{gammaitoni1998,
anishchenko1999} is 
based on appropriate measures. These
measures depend
in a non-monotonous way on
the noise intensity \cite{gammaitoni1998,anishchenko1999}. An increase of the noise intensity to a certain optimal level improves the output signal quality as measured by signal-to-noise ratio, spectral power amplification, residence time distribution \cite{gammaitoni1995,marchesoni2000}, probability of a given number of transitions per period of an external driving \cite{talkner2005}.
Stochastic resonance is not only a property of dynamical systems but it is also a property of level crossing triggered devices \cite{gingl1995,Wiesenfeld1994}  or time series sequences \cite{bezrukov1997}.

Usually it is assumed that noise  in physical systems is Gaussian. This is a direct consequence of the central limit theorem saying that a sum of independent bounded (characterized by a finite variance) random variables converges to the Gaussian distribution.
Nevertheless, the gathered experimental evidence suggests that there is a need to consider a more general type of noises. The presence of a more general, heavy-tailed fluctuations has been recorded in versatility of situations: diffusion in the energy space \cite{chechkin2002b}, exciton and charge transport in polymers under conformational motion \cite{lomholt2005}, spectral analysis of paleoclimatic \cite{ditlevsen1999b,ditlevsen2005} and economic data \cite{santini2000}, motion in optimal search strategies among randomly distributed target sites \cite{Viswanathan1996,sims2008,Edwards2007},  two-dimensional rotating flows \cite{solomon1993}. The area of applicability of L\'evy stable noises is steadily growing over time
including noise induced effects \cite{chechkin2006,metzler2007,dybiec2007e,dubkov2008,klages2008,dybiec2009}, epidemiology \cite{janssen1999,dybiec2008g}, ecology \cite{reynolds2009} and many others.


The current research extends earlier studies on the non-dynamical stochastic resonance induced by the Gaussian noise. Here, it is assumed that the noise is more general, i.e. the Gaussian noise is replaced by the non-equilibrium, heavy tailed $\alpha$-stable noise.  The findings presented in the
following sections extend existing studies on the role of L\'evy flights in physical systems.

\section{Model and results \label{sec:model}}

\begin{figure}%
\includegraphics[width=\columnwidth]{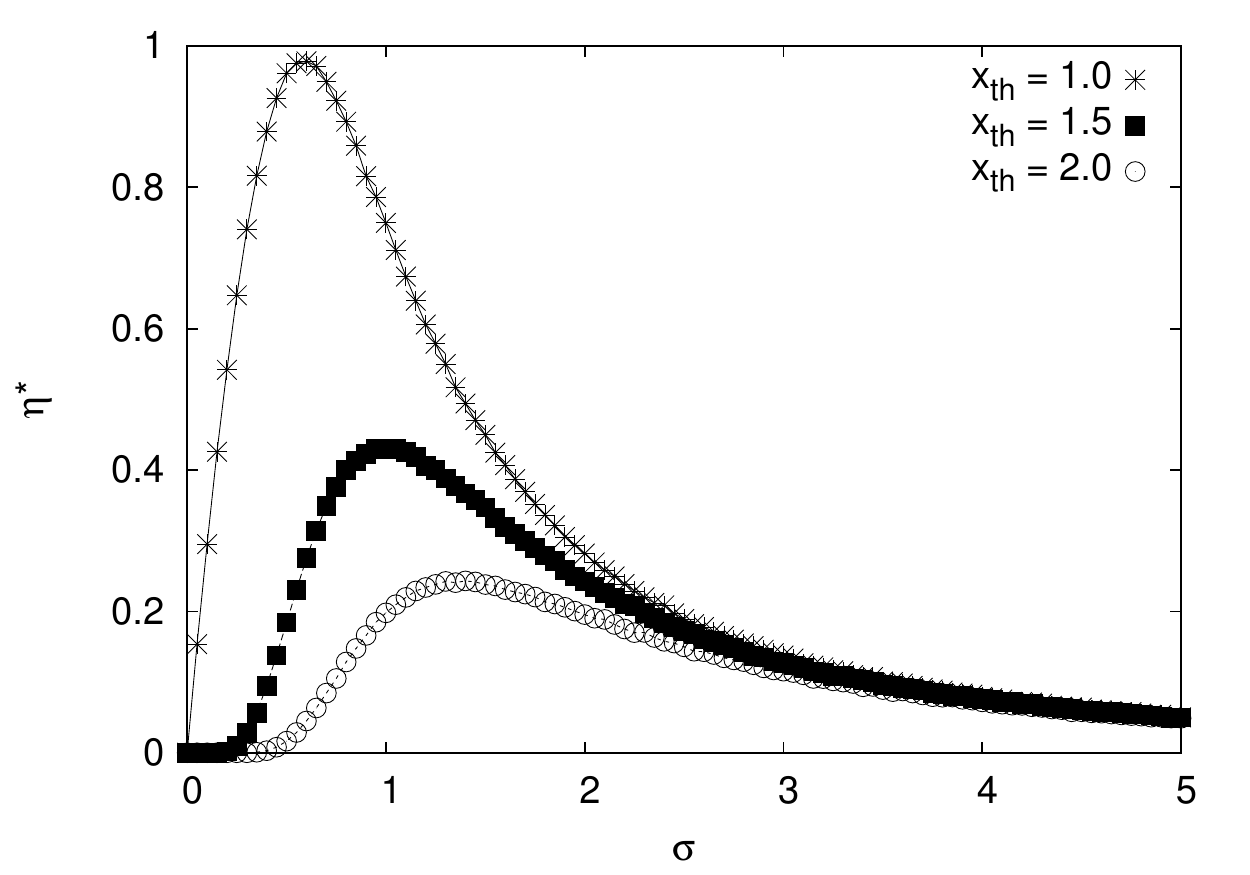}%
\caption{The rescaled spectral power amplification $\eta^*$  for the $\alpha$-stable white noise with $\alpha=2$ (stability index), $\beta=0$ (asymmetry parameter), i.e. for the white Gaussian noise, with various threshold $\xt=\{1,1.5,2\}$ as a function of the scale parameter $\sigma$.
To increase the readability of figures, recorded values of the spectral power amplification are divided by a normalization constant, which is fixed for all parameters in a figure.
}%
\label{fig:a20}%
\end{figure}

We study properties of the following system
\begin{equation}
 x(t)=\sin(\Omega t) + \sigma\zeta_{\alpha,\beta}(t),
\end{equation}
where $\zeta_{\alpha,\beta}(t)$ are independent $\alpha$-stable random variables distributed according to the $\alpha$-stable density $p_{\alpha,\beta}(x,\sigma=1,\mu=0)$, see next paragraph. The analyzed signal $y(t)$ is obtained by means of digital filtering of $x(t)$ as
\begin{equation}
 y(t)=
 \left\{
 \begin{array}{lcl}
  1 & \mbox{if} & x(t) > \xt \\
  0 & \mathrm{if} & x(t) \leqslant \xt \\
 \end{array}
 \right..
\end{equation}
Such a system is an archetypal model for the non-dynamical stochastic resonance \cite{bezrukov1997,gingl1995,Wiesenfeld1994}. Contrary to earlier examinations, here, it is assumed that the Gaussian noise is replaced by the more general $\alpha$-stable noise.

\begin{figure}%
\includegraphics[width=\columnwidth]{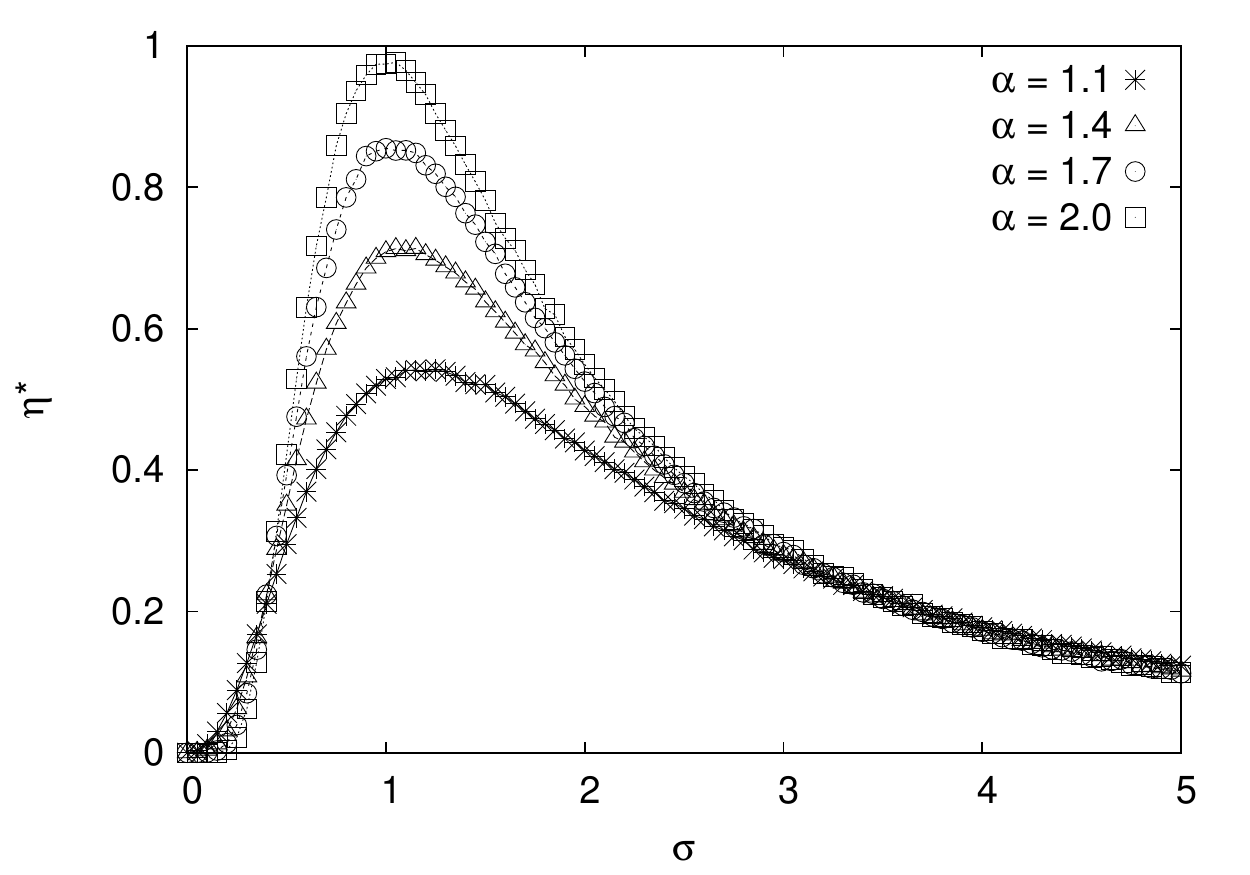}%
\caption{The rescaled spectral power amplification $\eta^*$  for symmetric $\alpha$-stable noises ($\beta=0$) with various stability index $\alpha=\{2,1.7,1.4,1.1\}$ as a function of the scale parameter $\sigma$. The threshold is set to $\xt=1.5$.}%
\label{fig:b00}%
\end{figure}

$\alpha$-stable random variables are distributed according to the density  function $p_{\alpha,\beta}(x;\sigma,\mu)$ with the characteristic function ($\phi(k)=\int_{-\infty}^\infty \mathrm{e}^{ikx}p_{\alpha,\beta}(x;\sigma,\mu)dx$)  \cite{janicki1994,janicki1996} 
\begin{equation}
\phi(k)=\left\{
\begin{array}{l}
 \exp\left[ -\sigma^\alpha|k|^\alpha\left( 1-i\beta\mathrm{sgn}(k)\tan\frac{\pi\alpha}{2} \right) +i\mu k \right] \\
 \;\;\;\;\;\;\mbox{for}\;\;  \alpha\neq 1, \\
 \exp\left[ -\sigma|k|\left( 1+i\beta\frac{2}{\pi}\mathrm{sgn} (k) \ln|k| \right) + i\mu k \right]\\
 \;\;\;\;\;\;\mbox{for}\;\;  \alpha= 1. \\
\end{array}
\right.
\end{equation}
Stable densities are characterized by four parameters: the stability index $\alpha$ ($\alpha\in(0,2]$), the asymmetry parameter $\beta$ ($\beta\in[-1,1]$), the scale parameter $\sigma$ ($\sigma>0$) and the location parameter $\mu$ ($\mu\in\mathbb{R}$). The stability index $\alpha$ describes asymptotic behavior of stable densities, i.e. for a large $x$ stable densities with $\alpha<2$ decay as a power law  $|x|^{-(\alpha+1)}$. The asymmetry parameter $\beta$ characterizes skewness of the distribution \cite{janicki1994,janicki1996}. For $\beta=0$ stable densities are symmetric ones while for $\beta \neq 0$ they are asymmetric. Finally, $\sigma$ describes the overall distribution width. In the limiting case of $\alpha=2$ the Gaussian density is recovered. In such a case, $\mu$ represents the mean value and $\sigma$ stands for the standard deviation. In the further studies it is assumed that $\mu=0$.

Depending on the threshold value $\xt$ and noise parameters, the recorded signal  $y(t)$ can display some periodicity. The pronounced periodicity is observed when noise pulses are strong enough to induce threshold crossing events only when $x(t)$ is close to the threshold, i.e. when $\sin(\Omega t) \approx 1$  (assuming that $\xt> 1$), see below. The periodicity of the recorded signal can be detected by the standard measures of the stochastic resonance: spectral power amplification or signal to noise ratio.

\begin{figure}%
\includegraphics[width=\columnwidth]{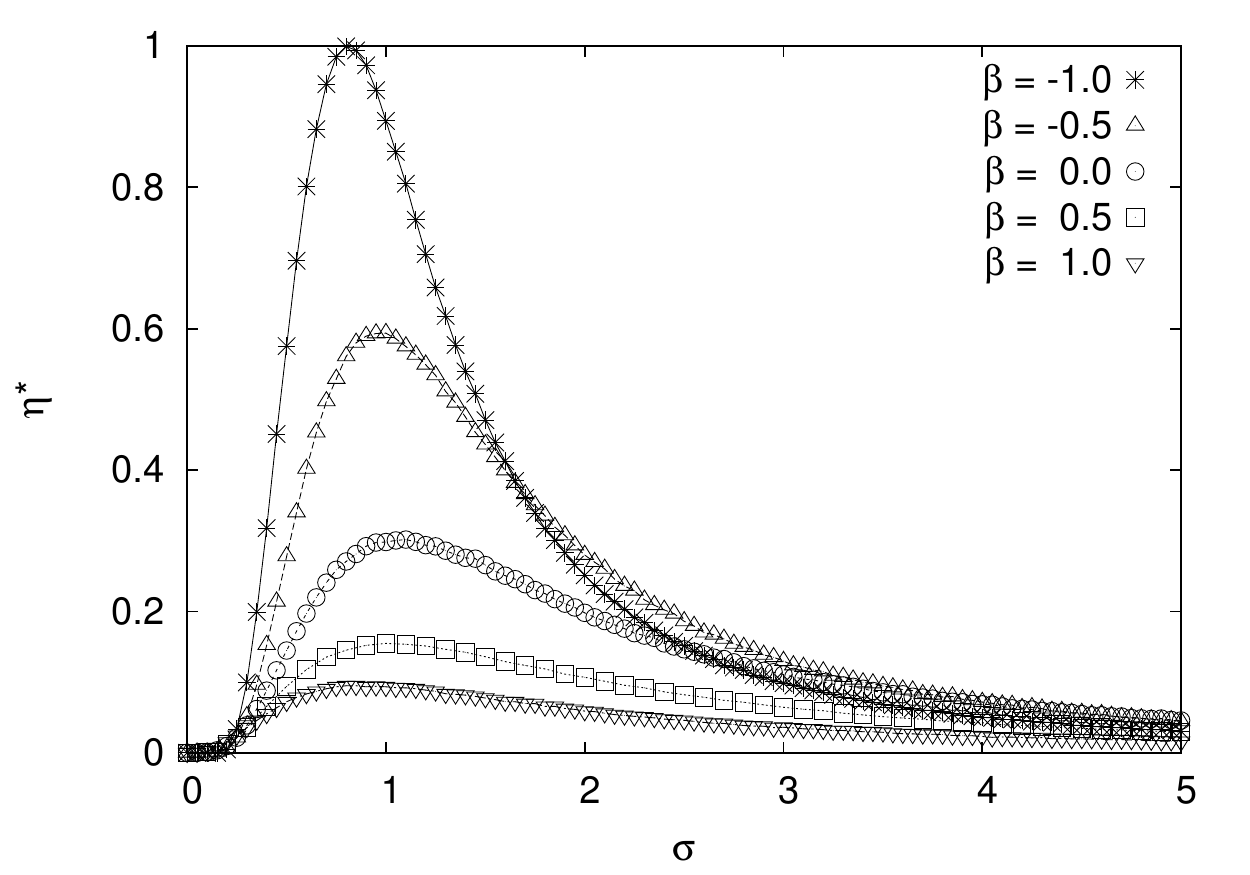}%
\caption{The rescaled spectral power amplification $\eta^*$  for $\alpha=1.5$ (stability index), $\xt=1.5$ (threshold) with various asymmetry parameter $\beta=\{-1,-0.5,0,0.5,1\}$ as a function of the scale parameter $\sigma$.}%
\label{fig:a15}%
\end{figure}

The spectral power amplification $\eta$ and the signal to noise ratio are the most common measures of the stochastic resonance \cite{gammaitoni1998}. Both of them are derived from power spectra $S(\omega)$. Assuming that there is a periodic input with an angular frequency $\Omega$ the spectral power amplification \cite{gammaitoni1998} is given by
\begin{equation}
\eta=\frac{p_1}{p_2},
\label{eq:spa}
\end{equation}
where $p_1$ is a power carried in delta-like spikes of $S(\omega)$ at the driving frequency $\Omega$, while $p_2$ is a power carried by the input signal. Therefore, the spectral power amplification measures relative amplification of the output at the driving frequency. The signal to noise ratio \cite{gammaitoni1998} measuring separation of the output from the noisy background is defined as
\begin{equation}
SNR=2\frac{\lim\limits_{\Delta \omega\to 0}\int_{\Omega-\Delta\omega}^{\Omega+\Delta\omega}S(\omega)d\omega}{S_N(\Omega)},
\label{eq:snr}
\end{equation}
where $S_N(\Omega)$ is a background level.

\begin{figure}%
\includegraphics[width=\columnwidth]{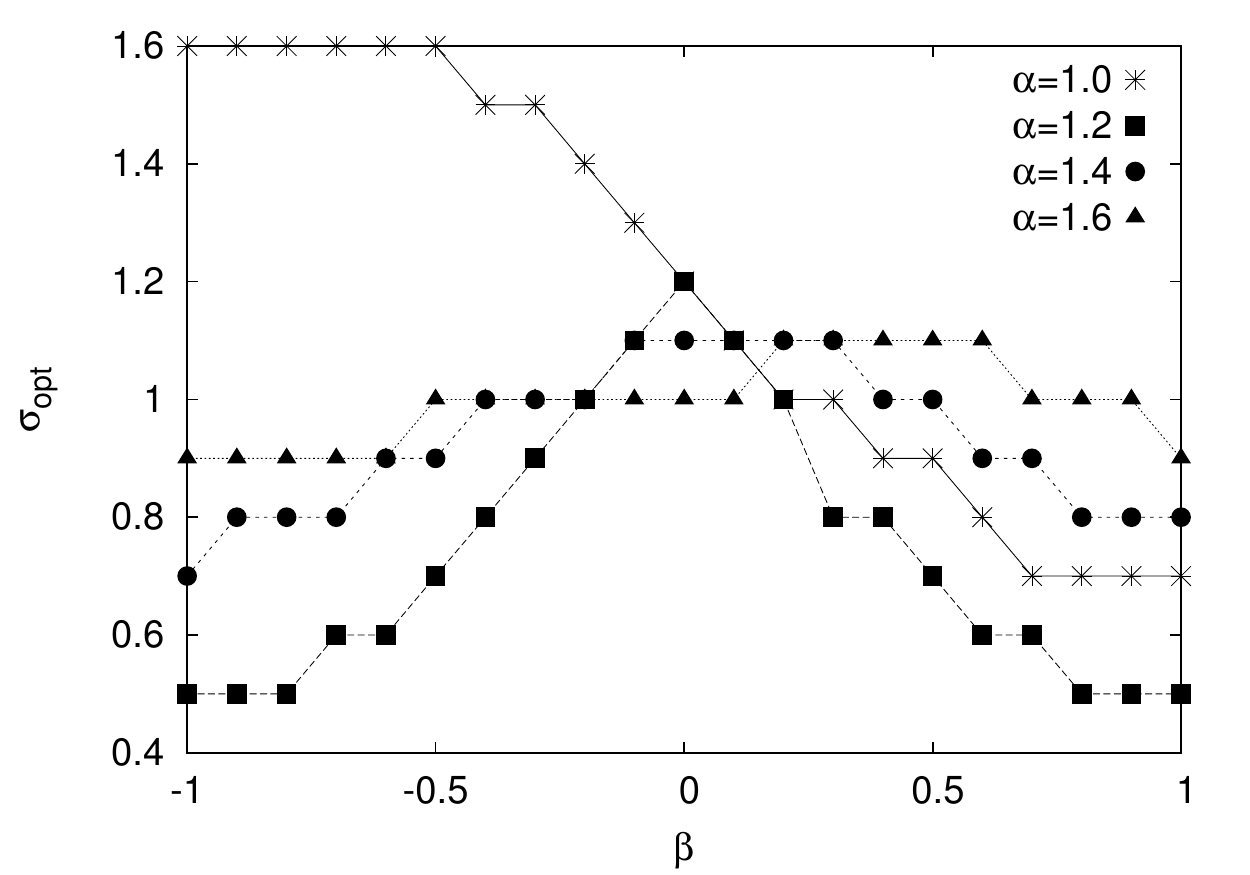}%
\caption{The optimal scale parameter $\so$, i.e. the scale parameter $\sigma$ leading to the largest values of the spectral power amplification, as a function of the asymmetry parameter $\beta$. The threshold level  is set to $\xt=1.5$. Various curves correspond to different values of the stability index $\alpha$.}%
\label{fig:optimalnoise}%
\end{figure}

\begin{figure}%
\includegraphics[width=\columnwidth]{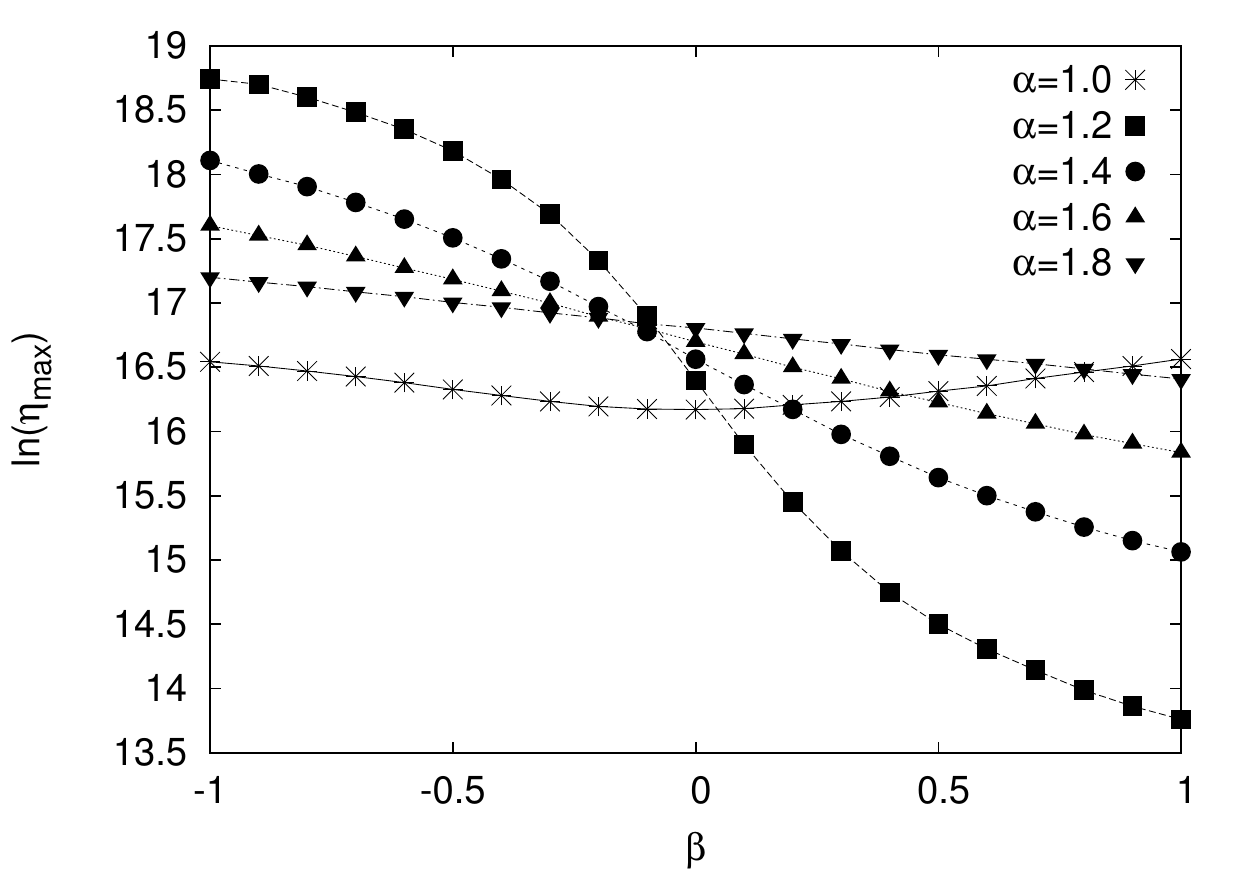}%
\caption{The maximal spectral amplification $\etamax$ as a function of the asymmetry parameter $\beta$. The threshold level  is set to $\xt=1.5$. Various curves correspond to different values of the stability index $\alpha$.}%
\label{fig:maxeta}%
\end{figure}

For $\alpha=2$ any $\alpha$-stable noise is equivalent to the Gaussian noise. Therefore, obtained results reproduce earlier findings on the non-dynamical stochastic resonance, see \cite{bezrukov1997,gingl1995,Wiesenfeld1994} and Fig.~\ref{fig:a20}. The signal to noise ratio and the spectral power amplification depend in non-monotonous way on the noise intensity $\sigma$. For a large enough threshold $\xt$ and a small noise intensity $\sigma$, the process $x(t)$ is always sub-threshold, i.e. $x(t) < \xt$. Consequently  $y(t) \equiv 0$ and no signal is recorded. With the increasing noise intensity, the process $x(t)$ can cross the threshold $\xt$. If the noise intensity is small, these crossings can take place only at time instants $t_i$ when $\sin(\Omega t_i)$ is maximal (closest to the threshold), i.e. when $\sin(\Omega t) \approx 1$. For even larger noise intensities the exact value of $\sin(\Omega t)$ is not important, because noise pulses are large enough to make $x(t)$ supra-threshold and $y(t)$ becomes 
insensitive to the periodic
modulation $\sin(\Omega t)$. The described mechanism explain non-monotonous dependence of stochastic resonance measures on the noise intensity, see Fig.~\ref{fig:a20}. Furthermore, the increase in the threshold level $\xt$ weakens the non-dynamical stochastic resonance and shifts slightly the optimal noise intensity towards larger values, see Fig.~\ref{fig:a20}. On the one hand, the increase in the optimal noise intensity is produced by an increasing gap between the maximal value of the periodic signal and the threshold. On the other hand, the increase in the noise intensity decreases the strength of resonance.

The $\alpha$-stable noise is characterized not only by the scale parameter $\sigma$ but also by the stability index $\alpha$. Both these parameters control the distribution width as measured by the interquantile distance (note that $\alpha$-stable distributions are characterized by the infinite variance). For $\alpha=2$, $\alpha$-stable densities are equivalent to the Gaussian density.
Fig.~\ref{fig:b00} presents the spectral power amplification $\eta$ as a function of the scale parameter $\sigma$ for symmetric ($\beta=0$) $\alpha$-stable noises. Various curves correspond to various values of the stability index  $\alpha$. The threshold $\xt$ is set to $\xt=1.5$. In comparison to the Gaussian case ($\alpha=2$), in the non-equilibrium regime, the optimal scale parameter $\so$ shifts insignificantly towards larger values with the decreasing value of the stability index $\alpha$, see Figs.~\ref{fig:b00} and~\ref{fig:optimalnoise}. Moreover, the maximal values of spectral power amplification $\eta$ for symmetric noises ($\beta=0$) significantly decay with the decrease of the stability index $\alpha$, see Fig.~\ref{fig:b00}. The decay of the maximal spectral power amplification originates in the increase of the distribution width with the decrease of the stability index $\alpha$. Consequently, in this situation the decrease in $\alpha$ acts in the same manner like the increase in the scale 
parameter $\sigma$, which also 
leads to the decay of the spectral power amplification.

\begin{figure}[!h]%
\includegraphics[width=\columnwidth]{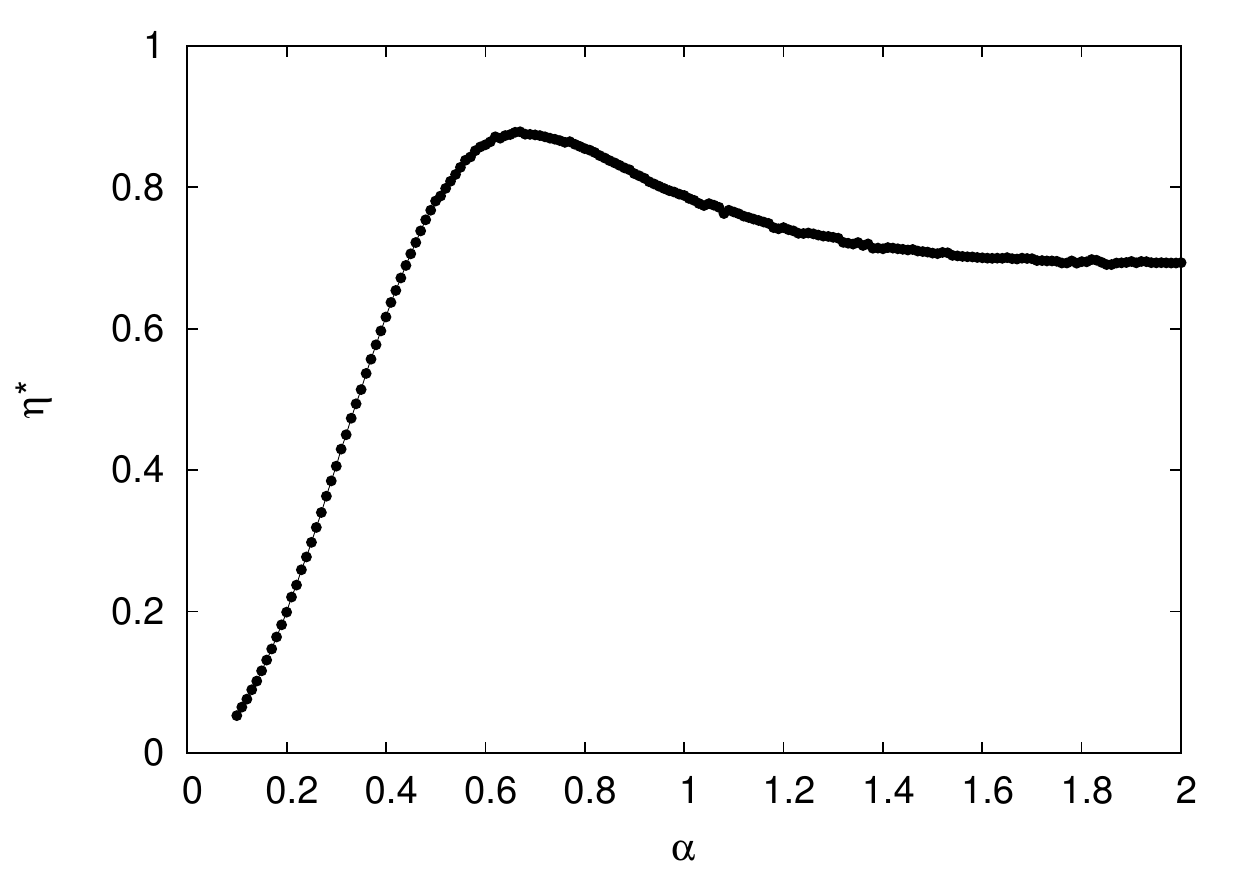}\\%
\includegraphics[width=\columnwidth]{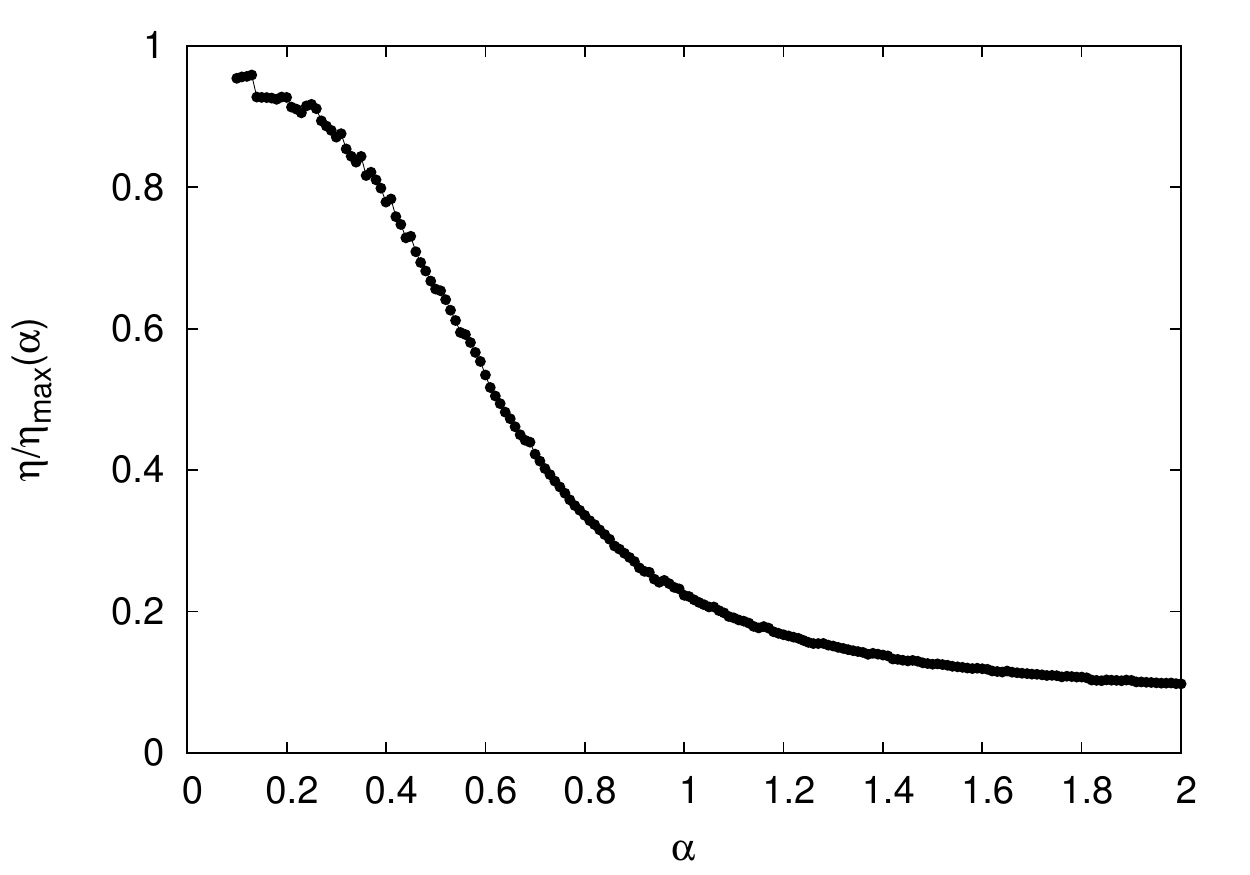}%
\caption{The rescaled spectral power amplification $\eta^*$ for $\xt=1.5$ (threshold), $\beta=0$ (asymmetry parameter), $\sigma=5.5$ (scale parameter) as a function of the stability index $\alpha$ (top panel) and the spectral power amplification $\eta$ divided by the maximal spectral power amplification $\eta_{\mathrm{max}}(\alpha)$ (bottom panel).
}%
\label{fig:sigma055}%
\end{figure}

The asymmetry of the noise also affects the model performance. Fig.~\ref{fig:a15} presents the spectral power amplification as a function of the scale parameter $\sigma$ for various asymmetry parameters $\beta$. The stability index $\alpha$ is set to $\alpha=1.5$ while the threshold $\xt$ is $\xt=1.5$. $\alpha$-stable densities with the stability index $\alpha<1$ and the asymmetry parameter $\beta=\pm1$ are fully skewed, i.e. for $\beta=-1$ random numbers distributed according to these densities are always smaller than location parameter $\mu$, which is set to $\mu=0$, while for $\beta=1$ they are always larger than $\mu$. Consequently, when $|\xt|>1=\mathrm{max}(\sin(\Omega t))$ the process $x(t)$ can be always sub-threshold, because playing with noise parameters it is possible to produce noise pulses which shift $x(t)$ towards negative values and consequently makes $y(t)\equiv 0$. For example,  such a situation is observed for $\xt=1.5$ with $\alpha<1$ and $\beta=-1$ and any value of the scale
parameter $\sigma$. 
Contrary to the extreme
case ($\alpha<1$ and $\beta=\pm 1$) the changes in the asymmetry parameter lead to richer behavior of the spectral power amplification, see below.
For $\alpha>1$, the increase in the asymmetry parameter results in weakening of the non-dynamical stochastic resonance.
With increasing $\beta$ heights of recorded resonance curves decrease. At the same time width of resonance curves increase, see Fig.~\ref{fig:a15}, what is further confirmed in the bottom panel of Fig.~\ref{fig:sigma055}.

Figure~\ref{fig:optimalnoise} presents the optimal scale parameter $\so$ as a function of the asymmetry parameter $\beta$. Various curves correspond to different values of the stability index $\alpha$. When the stability index $\alpha$ is close to 2, i.e. the noise is close to the Gaussian noise, the optimal scale parameter $\so$ displays very weak sensitivity or lack of sensitivity to the asymmetry parameter $\beta$ because changes in $\beta$ induce only minor changes in the shape of $\alpha$-stable densities.
Complementary Fig.~\ref{fig:maxeta} presents the spectral power amplification corresponding to the optimal noise intensity $\so$, i.e. maximal values of the spectral amplification $\etamax$,  as a function of the asymmetry parameter $\beta$. Various curves correspond to different values of the stability index $\alpha$. For symmetric noise ($\beta=0$), the decrease in the stability index $\alpha$ weakens the strength of the non-dynamical stochastic resonance (as documented in Fig.~\ref{fig:b00}). The very different situation is observed for asymmetric noises ($\beta \neq 0$), when the non-dynamical stochastic resonance can be either weakened or enhanced in comparison to the reference Gaussian case. For the fixed $\alpha$ with the decreasing value of the asymmetry parameter $\beta$ left tail of the $\alpha$-stable density becomes heavier and heavier. At the same time (for $\alpha>1$) the modal value moves to the right. Therefore, a smaller scale parameter $\so$ leads to the maximal spectral power amplification 
$\etamax$. The decrease in 
the asymmetry parameter leads to the increase in the maximal spectral power amplification $\etamax$.  When $\alpha$ approaches 2 changes in $\etamax$ are small, because changes in the noise distribution are minimal.

Traditionally the system is tuned to the stochastic resonance by adjusting the scale parameter $\sigma$ (noise intensity). Since $\alpha$-stable noises are characterized by four parameters it is possible to fine tune the system by changing values of the stability index $\alpha$ or the asymmetry parameter $\beta$, see Fig.~\ref{fig:maxeta}. Top panel of Fig.~\ref{fig:sigma055} presents a cross section through a spectral amplification surface at a fixed value of the scale parameter $\sigma=5.5$ for symmetric $\alpha$-stable noises. The threshold level $\xt$ is set to $\xt=1.5$. The well pronounced maximum of spectral power amplification is recorded at $\alpha \approx 0.7$. The bottom panel of Fig.~\ref{fig:sigma055} shows fraction of the maximal spectral power amplification observed for a given value of the stability index $\alpha$, i.e. $\eta(\alpha,\beta=0,\sigma=5.5)/\eta_{\mathrm{max}}(\alpha,\beta=0)$. The fraction of the maximal spectral power amplification is decreasing function of the stability index $\
alpha$ 
because with decreasing $\alpha$ resonance curves flatten, see Fig.~\ref{fig:a15}.

\section{Summary and conclusions\label{sec:summary}}

An $\alpha$-stable noise provides natural generalization of the Gaussian noise. The generalized central limit theorem together with well deweloped numerical methods makes $\alpha$-stable noises especially suitable for approximation of far-from-equilibrium fluctuations.

Analogously, like in the case of equilibrium fluctuations, non-equilibrium heavy tailed $\alpha$-stable noises can induce the non-dynamical stochastic resonance. In comparison to the Gaussian non-dynamical stochastic resonance, the strength of recorded resonances can be significantly enhanced or weakened by $\alpha$-stable noise. In the close to Gaussian regime ($\alpha \lessapprox 2$) the system performance display weak sensitivity to stable noise parameters. The largest sensitivity to the exact shape of noise pulses is observed in the far from Gaussian regime, especially for asymmetric noises. The strength of the non-dynamical stochastic resonance is not only controlled by the scale parameter (noise intensity) but also by remaining noise parameters: stability index and asymmetry parameter.


\begin{acknowledgments}
Computer simulations have been performed at the Academic
Computer Center Cyfronet, Akademia G\'orniczo-Hutnicza (Krak\'ow, Poland) under CPU grant
MNiSW/Zeus\_lokalnie/UJ/052/2012.

\end{acknowledgments}


\end{document}